\documentclass[a4paper]{jpconf}
\usepackage{graphicx}
\begin{document}
\title{Novel theoretical approach in photoemission spectroscopy: application
to isotope effect and boron-doped diamond}

\author{Jifeng Yu, Kai Ji and Keiichiro Nasu}

\address{
CREST JST, Solid State Theory Division,
Institute of Materials Structure Science, KEK,
Graduate University for Advanced Studies,
Oho 1-1, Tsukuba, Ibaraki 305-0801, Japan
}

\ead{jikai@post.kek.jp}

\begin{abstract}
A new path-integral theory is developed to calculate the photoemission
spectra (PES) of correlated many-electron systems.
The application to the study on Bi$_2$Sr$_2$CaCu$_2$O$_8$ (Bi2212) and
boron-doped diamond (BDD) is discussed in details.
It is found that the isotopic shift in the angle-resolved photoemission
spectra of Bi2212 is due to the off-diagonal quadratic electron-phonon
($e$-ph) coupling, whereas the presence of electron-electron repulsion
partially suppresses this effect.
For the BDD, a semiconductor-metal phase transition, which is induced
by increasing the $e$-ph coupling and dopant concentration, is reproduced
by our theory.
Additionally, the presence of Fermi edge and phonon step-like structure
in PES is found to be due to a co-existence of itinerant and localized
electronic states in BDD.
\end{abstract}

\section{Introduction}

The role of electron-phonon ($e$-ph) interaction in the high-$T_c$
superconductivity has received considerable attention since the discovery
of oxygen isotope effect of Bi$_2$Sr$_2$CaCu$_2$O$_8$ (Bi2212) in the
angle-resolved photoemission spectra (ARPES)\cite{gw04,do07}.
The experimental data show that electronic band is shifted slightly with
the $^{16}$O/$^{18}$O substitution, signifying the existence of $e$-ph
interplay in cuprates.
Besides, theoretically clarifying this effect is of great significance,
for the energy scale of shifts reported by the two groups in Refs. \cite{gw04}
and \cite{do07} seem quite inconsistent with each other, and many questions
still remain up to now.
In order to have an insight into the isotope effect, in this work, we
develop a new path-integral theory to calculate the photoemission spectra
(PES) of cuprate superconductors, in which the electron-electron ($e$-$e$)
and $e$-ph correlations are treated on an equal footing.
This theory is also applicable to other kind correlated materials.
As an example, here, we also study the PES of boron-doped diamond (BDD),
which undertakes a semiconductor-metal phase transition on doping, and
becomes a superconductor with the temperature decreases\cite{ek04}.
The details of our theory will be presented in the next section, in connection
with the study on isotope effect.
Calculation and discussion on PES of BDD are included in Section 3.
A concluding remark can be found in the Summary.

\section{Isotopic shift in ARPES of Bi$_2$Sr$_2$CaCu$_2$O$_8$}

\subsection{Model for CuO$_2$ plane of cuprate superconductor}

In the CuO$_2$ plane of cuprates, the electronic transfer is modulated
by the vibration of oxygen atoms between the initial and final Cu sites
(see in Fig. 1), resulting in an off-diagonal type $e$-ph coupling.
In order to qualitatively clarify the isotope effect of Bi2212, we start
from a half-filled Hamiltonian including the $e$-$e$ repulsion and the
above mentioned off-diagonal $e$-ph coupling
($\hbar = 1$ and $k_B = 1$ throughout this paper):
\begin{eqnarray}
H &=& - \sum_{\langle l,l' \rangle, \sigma} t (l, l')
  (a^{\dag}_{l \sigma} a_{l' \sigma} + a^{\dag}_{l' \sigma} a_{l \sigma})
  +U \sum_l n_{l \uparrow} n_{l \downarrow}
  + {\omega_0 \over 2} \sum_{\langle l,l' \rangle} \left(-{1 \over \lambda}
  \frac{\partial^2}{\partial q^2_{ll'}} + q^2_{ll'} \right),
\end{eqnarray}
where $a^{\dag}_{l \sigma}$ ($a_{l \sigma}$) is the creation (annihilation)
operator of an electron with spin $\sigma$ at the Cu site $l$ on a square
lattice (Fig. 1).
The electrons hop between two nearest neighboring Cu sites, denoted by
$\langle l, l' \rangle$, with a transfer energy $t(l, l')$.
$U$ is the strength of Coulomb repulsion between two electrons on the
same Cu site with opposite spins.
The oxygen phonon is assumed to be of the Einstein type with a frequency
$\omega_0$ and a mass $m$.
$\lambda$ ($\equiv 1 + \Delta m / m$)
is the mass change factor of phonon due to the isotope substitution.
In the third term, $q_{ll'}$ is the dimensionless coordinate operator
of the oxygen phonon locating between the nearest-neighboring Cu sites
$l$ and $l'$, and the sum denoted by ${\langle l,l' \rangle}$ just means
a summation over all the phonon sites in the lattice.

\begin{figure}[h] 
\begin{center}
\includegraphics[width=8pc]{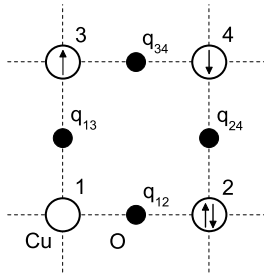}
\hspace{2pc}
\begin{minipage}[b]{20pc}
\caption{Lattice structure of CuO$_2$ conduction plane in cuprates.
The copper atom (white circle) is on the $l$-th site of a simple square lattice,
where the electrons (arrows) reside.
The oxygen atom (black circle) is located between the nearest-neighboring Cu
sites, and $q_{l l'}$ denotes its displacement from the equilibrium position.}
\end{minipage} 
\end{center}
\end{figure}

In the conduction plane of CuO$_2$, the electronic hopping integral
$t (l, l')$ can be expanded to the second order terms with respect to
the phonon displacements $q_{ll'}$ as
\begin{equation}
t (l, l') = t_0 + s q^2_{ll'},
\end{equation}
where $t_0$ is the bare hopping energy and $s$ the off-diagonal quadratic
$e$-ph coupling constant.
Here we note the linear $e$-ph coupling does not occur owing to the lattice
symmetry of present model.
Whereas the inter-site $e$-$e$ interaction is included in the screened
values of $t_0$ and $s$.

\subsection{Path-integral Monte Carlo method}

In this section, we develop a path-integral theory for a model with both
$e$-$e$ and $e$-ph interactions.
By making use of the Trotter's decoupling formula, the Boltzmann operator
is written as,
\begin{eqnarray}
e^{-\beta H} = \lim_{L \rightarrow \infty}
	\left( e^{-\Delta H} \cdots e^{-\Delta H} \right).
\end{eqnarray}
Applying the Hubbard-Stratonovitch transformation\cite{to97} and the Gaussian
integral formula\cite{ji04}, we can decouple the two-body parts, so that the
$e$-$e$ and $e$-ph correlated terms are replaced by a two-fold summation
over the auxiliary spin and lattice configurations, which is the so-called
path-integral.
In this way, the Boltzmann operator is rewritten into the path-integral
form as,
\begin{eqnarray}
e^{-\beta H} & \rightarrow & \int {\mathcal D} x
  \left( T_+ \exp \left\{ -\int_0^\beta d \tau
  \left[ h(\tau,x_m,x_q) + \Omega(x_q) \right] \right\}
  \prod_l \left[ |x_q(l,\beta) \rangle \langle x_q(l,0)|
  \right] \right),
\end{eqnarray}
\begin{eqnarray}
h(\tau,x_m,x_q) & \equiv & - \sum_{\langle l,l' \rangle, \sigma}
  \left[ t_0 + s x^2_q(l,l',\tau) \right]
  \left[a^{\dag}_{l \sigma}(\tau) a_{l' \sigma}(\tau)
  + a^{\dag}_{l' \sigma}(\tau) a_{l \sigma}(\tau) \right]
  \nonumber\\
& &	- \sqrt{U \over \Delta} \sum_l x_m(l, \tau)
  [ n_{l \uparrow}(\tau) - n_{l \downarrow}(\tau) ],
  \\
  \Omega(x_q) & \equiv & \sum_{\langle l,l' \rangle}
  \left\{ {\lambda \over 2 \omega_0}
  \left[ {\partial x_q(l,l',\tau) \over \partial \tau} \right]^2
  + {1 \over 2} \omega_0 x^2_q(l,l',\tau) \right\}.
\end{eqnarray}
Here, $x_m$ and $x_q$ correspond to the auxiliary spin and lattice field,
respectively, $\int {\mathcal D} x$ symbolically denotes the integrals
over the path $x$ synthesized by $x_m$ and $x_q$, and $|x_q \rangle$ is
the eigenstate of phonon.
$\Delta$ is the time interval of the Trotter's formula,
$\beta \equiv 1/T$, and $T$ is the absolute temperature.
$T_+$ in Eq. (4) is the time ordering operator.

Then the time evolution operator [$\equiv R (\tau, x)$] along a path $x$
is defined as
\begin{eqnarray}
R(\tau,x) = T_+ \exp \left[ -\int_0^{\tau} d \tau'
	h ( \tau', x_m, x_q ) \right].
\end{eqnarray}
In terms of the Boltzmann operator (4) and time evolution operator (7),
we define the free energy [$\equiv \Phi (x)$] of the given path as
\begin{eqnarray}
e^{-\beta \Phi (x)} = e^{-\int_0^{\beta} d \tau \Omega(x_q)}
	{\rm Tr} \left[ R(\beta, x) \right].
\end{eqnarray}
While, the partition function ($\equiv Z$) and total free energy
($\equiv \Phi$) are given as
\begin{eqnarray}
Z = e^{-\beta \Phi} = \int {\mathcal D}x
  e^{-\beta \Phi (x)}.
\end{eqnarray}
According to Refs. \cite{to97} and \cite{ji04}, we also define the one-body
Green's function [$\equiv G_{\sigma} (l \tau, l' \tau', x)$] on a path $x$ as
\begin{eqnarray}
G_{\sigma} (l \tau, l' \tau', x) = - \mbox{sign} (\tau - \tau')
	\langle T_+ \vec{a}_{l \sigma} (\tau)
	\vec{a}^{\dag}_{l' \sigma} (\tau') \rangle_x,
\end{eqnarray}
where $\vec{a}_{l \sigma} (\tau)$ is the Heisenberg representation of
$a_{l \sigma}$.
It is really time-dependent and defined by
\begin{eqnarray}
\vec{a}_{l \sigma} (\tau) \equiv R^{-1}(\tau,x) a_{l \sigma} R(\tau,x) .
\end{eqnarray}
Meanwhile, the ordinary Green's function [$\equiv G_{\sigma} (l, \tau)$]
can be obtained by the path-integral as
\begin{equation}
G_{\sigma} (l - l', \tau - \tau') = {1 \over Z} \int
	{\mathcal D}x e^{- \beta \Phi (x)}
	G_{\sigma} (l \tau, l' \tau', x).
\end{equation}
This path-integral is evaluated by the quantum Monte Carlo (QMC) simulation
method.

If the QMC data of Green's function $G_{\sigma} (l, \tau)$ is obtained,
we can immediately calculate its Fourier component
[$\equiv G_{\sigma} ({\bf k}, \tau)$] as
\begin{eqnarray}
G_{\sigma} ({\bf k}, \tau) = {1 \over N} \sum_l
	G_{\sigma} (l, \tau) e^{-i {\bf k} \cdot {\bf R}_l} ,
\end{eqnarray}
where $\bf k$ is the momentum of the outgoing photo-electron.
From this Fourier component $G_{\sigma} ({\bf k}, \tau)$, we derive the
momentum-specified spectral function [$\equiv A_{\sigma} ({\bf k}, \omega)$]
by solving the integral equation
\begin{eqnarray}
G_{\sigma} ({\bf k}, \tau) = - \int^{\infty}_{- \infty} d \omega
	\frac {e^{- \tau \omega}} {1 + e^{- \beta \omega}}
	A_{\sigma} ({\bf k}, \omega) .
\end{eqnarray}

\subsection{Isotope substitution induced band shift in ARPES}

We now present the QMC results on a 4$\times$4 square lattice, where $t_0$
is set as the unit of energy, and $\omega_0$=1.0 is used.
For the QMC simulation, we impose a little large isotopic mass enhancement,
$\lambda_0$=1 and $\lambda$=2, to suppress the numerical error.
In this calculation, we determine the binding energy $\epsilon_{\bf k}$
by the moment analysis of the spectral function as $\epsilon_{\bf k} =
\sum_{\sigma} \int d \omega A_{\sigma}({\bf k}, \omega) \omega$.
Correspondingly, the isotope induced band shift is calculated by
$\Delta \epsilon_{\bf k} \equiv
\epsilon_{\bf k} (\lambda) - \epsilon_{\bf k} (\lambda_0)$.

\begin{figure}[h] 
\begin{center}
\includegraphics[width=26pc]{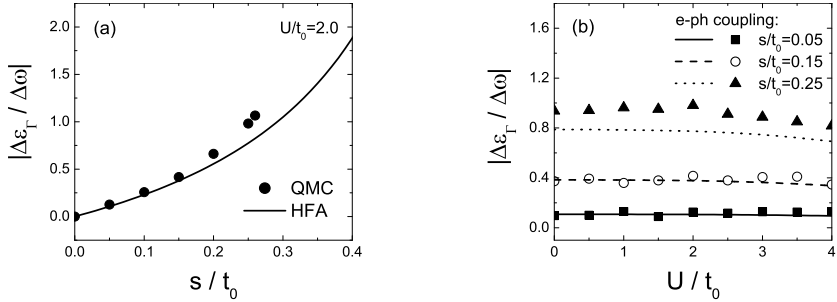}
\begin{minipage}{36pc}
\caption{
(a) The variation of $\Delta \epsilon_{\Gamma} / \Delta \omega$ with $s$
on a 4$\times$4 square lattice, when $U$=2.0, $\beta$=10, $\lambda_0$=1,
$\lambda$=2.
The filled circles are from QMC, and the solid curve from HFA as a guide
for eyes.
(b) The variations of $\Delta \epsilon_{\Gamma} / \Delta \omega$ with $U$
on a 4$\times$4 square lattice at $\beta$=10, $\lambda_0$=1 and $\lambda$=2.
Three different values of $s$ are used to show the $s$-dependence of
$\Delta \epsilon_{\Gamma} / \Delta \omega$.
The discrete symbols are the results of QMC, and continuous curves by
HFA as a reference.
}
\label{f2}
\end{minipage} 
\end{center}
\end{figure}

In Fig. 2(a), we plot the ratio $\Delta \epsilon_{\Gamma} / \Delta \omega$
versus $s$, at $U$=2.0 and $\beta$=10, where $\Delta \epsilon_{\Gamma}$ is
the band shift at the $\Gamma$ point of Brillouin zone
[${\bf k}_{\Gamma}$=(0,0)], and $\Delta \omega$ is the isotopic change
of phonon energy.
The filled circles are calculated by QMC, and the solid curve by the
mean-field theory with Hartree-Fork approximation (HFA) as a guide for
eyes.
Here both theories figure out an increase of
$\Delta \epsilon_{\Gamma} / \Delta \omega$ with $s$, which means if the
$e$-ph coupling is strong enough, a large band shift can be generated
in a small cost of $\Delta \omega$.
In Fig. 2(b), the ratio $\Delta \epsilon_{\Gamma} / \Delta \omega$ versus
$U$ are shown for three different $s$'s, where the discrete symbols and
continuous curves are the QMC and HFA results, respectively.
One can see the ratio $\Delta \epsilon_{\Gamma} / \Delta \omega$ increases
with $s$.
Meanwhile, for a fixed $s$, the ratio declines slightly as $U$ increases,
indicating that the band shift is owing to the $e$-ph coupling, whereas
the presence of $U$ partially reduces this effect.
According to Fig. 2, the band shift thus can be regarded as a measure
of the effective $e$-ph coupling strength in the system.

%
%

\section{Co-existence of localization and itineracy of electrons in
boron-doped diamond}

It is well known that the pristine diamond is a big band gap insulator.
Lightly doped with boron, it shows a $p$-type character with an activation
energy about 0.37 eV\cite{co71}.
Recently, the research on BDD has become highly attractive since the remarkable
discovery of superconductivity in this material\cite{ek04}.
Accompanied with the superconducting phase, a semiconductor-metal transition
also occurs when the doping percentage is increased to certain level.
In the normal metallic state, K. Ishizaka {\it et al}.\cite{is06} declared
the observation of a step-like multi-phonon satellite structure in the
valence band PES, approximately distributed periodically in 0.150 eV
below the Fermi level, in addition to the emergence of a clear Fermi edge.
This periodic structure in PES reminds us of a similarity to the case
of localized electron\cite{ma00}, wherein the coupling between electron
and Einstein phonons characterizes the spectra with discrete peaks at
equal distance.
Moreover, the Fermi edge and the step-like structure are observed together,
probably originating from a co-existence of the two basic properties
of electrons: itineracy and localization.
In order to clarify this so-called co-existence theoretically, we apply
our path-integral theory to the many-impurity-Holstein (MIH) model on
a doped simple cubic lattice to derive its spectral density.

\subsection{Model and methods}

The MIH model includes the following two properties.
One is the disorder of the system, that some atoms are replaced by dopant
ones in a certain ratio.
The other is the coupling between electrons and Einstein phonons, being
the simplest description of $e$-ph interactions.
Its Hamiltonian is given as,
\begin{eqnarray}
H&=&-t \sum_{\langle l, l'\rangle} \sum_{\sigma}
    (a^{\dag}_{l\sigma} a_{l'\sigma} + a^{\dag}_{l'\sigma} a_{l\sigma})
    -\mu \sum_{l, \sigma} n_{l \sigma}
    +\Delta_{e} \sum_{l_{0}, \sigma} n_{l_{0} \sigma}
    +\frac{\omega_{0}}{2} \sum_{l}
    \left(-\frac{{\partial}^2}{\partial Q_{l}^2} + Q_{l}^2 \right)
    \nonumber\\
& & -S \sum_{l,\sigma} Q_{l} (n_{l \sigma} - \overline{n}_{l}/2),
    \quad n_{l\sigma}\equiv a^{\dag}_{l\sigma}a_{l\sigma}, \;
    \sigma=\alpha \; \textrm{or} \; \beta,
\end{eqnarray}
where $t$ is the transfer energy. $a^{\dag}_{l\sigma}$ and
$a_{l\sigma}$ are the creation and annihilation operators of
electron with spin $\sigma$ at site $l$. Electrons can hop only
between the nearest neighboring sites expressed by $\langle
l,l'\rangle$. $\mu$ stands for the chemical potential of electrons
and  $\Delta_{e}$ is the potential difference between after and
before substitution at the doped sites labeled by $l_{0}$. $Q_{l}$
is the dimensionless coordinate operator for the phonon at site $l$
with frequency $\omega_{0}$. $S$ denotes the \emph{e}-ph coupling
constant. $\overline{n}_{l}$ is the average electron number at site
$l$. To simplify the problem, we just consider the coupling at the
doped sites hereafter, because in the pure diamond there is no
evidence that the satellite structure appears, which suggests
the coupling is important only after doping.

By applying our path-integral theory to the disordered system, and following
the formulation in Section 2.2, we obtain an electronic Green's function
similar to Eq. 12,
\begin{equation}
G_{\sigma} (l, \tau) = {1 \over Z} \int
	{\mathcal D}x e^{- \beta \Phi (x)}
	G_{\sigma} (l \tau, l 0, x).
\end{equation}
In the numerical calculation, the path-integral of Green's function in
Eq. 16 is also performed by the QMC simulation method as before.
After averaging this site-dependent Green's function over all the $N$
sites of the system, the photoemission spectral function
[$\equiv N_{\sigma}(\omega)$] can be reproduced through the analytic
continuation as,
\begin{equation}
\frac{1}{N} \sum_{l} G_{\sigma} (l, \tau) =
  -\int_{-\infty}^{+\infty}
  \frac{e^{-\tau \omega}}{1+e^{-\beta \omega}}
  N_{\sigma} (\omega) d \omega .
\end{equation}
Finally, after imposing the Fermi-Dirac function
$f (\omega) = 1 / [\exp(\beta \omega) + 1]$, the PES intensity is given
as $I(\omega) = \sum_{\sigma} N_{\sigma} (\omega) f(\omega)$,
which can be compared with the experimental data.

\subsection{Results and discussions}

Since we focus on the spectral region close to the Fermi level, we use
a simple cubic lattice of $4\times4\times4$ in real calculation without
paying much attention to the detail of carbon valence band.
As we just count the $e$-ph coupling at the doped sites, the phonon effect
is not obvious in the whole system PES after averaging over all sites.
For this reason, in the following, we will also present the PES of doped
sites to illustrate the phonon effect.

\begin{figure}[h]
\begin{minipage}{18pc}
\includegraphics[width=17pc]{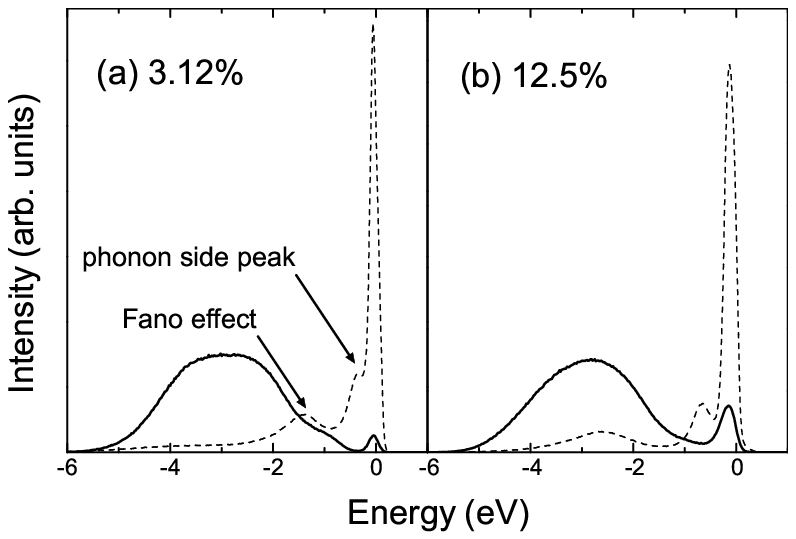}
\caption{
The QMC-calculated spectra for a simple cubic lattice of light (left)
and heavy (right) doping rates, at a weak $e$-ph coupling.
Zero is the position of Fermi level.
Full lines denote the PES of whole system, and dotted lines the spectra
of doped sites.
}
\end{minipage}\hspace{2pc}%
\begin{minipage}{18pc}
\includegraphics[width=14pc]{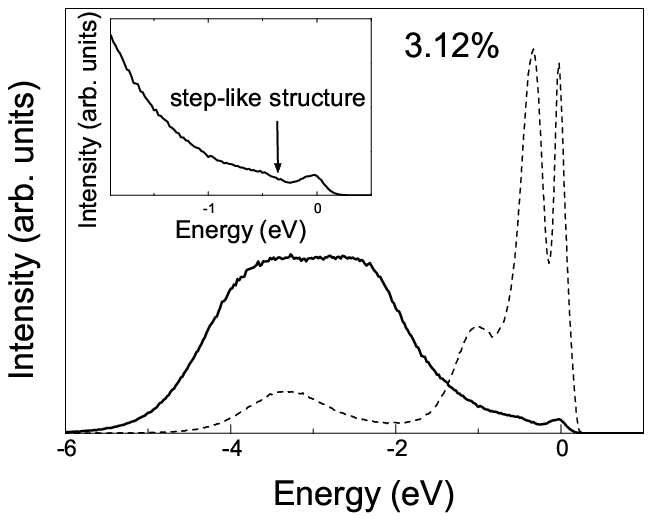}
\caption{
The co-existence of Fermi edge and step-like structure in the PES, at
a low doping rate and a strong $e$-ph coupling, by QMC.
Full (dotted) line is the spectrum of whole system (doped sites).
The inset zooms in the solid line near the Fermi level.
}
\end{minipage}
\end{figure}

Fig. 3 shows the spectra for the simple cubic lattice at light (3.12\%)
and heavy (12.5\%) doping rates, with a weak $e$-ph coupling of $S$=0.25 eV.
From the whole system PES (full lines), one can clearly see the emergence
of a Fermi edge with the increase of doping percentage.
In the lightly doped case, the impurity levels are a little above the
top of valence band, indicating the system is a semiconductor.
While in the heavily doped case, the impurity band expands to overlap
the top of valence band and closes the activation gap, leading to a
semiconductor-metal phase transition.
In the spectra of boron-doped sites (dotted lines), one notices that,
in addition to a broad hump due to the so-called Fano effect\cite{fa61},
a satellite structure also appears in each case a little below the Fermi level.
We should note this side structure is just the quantum phonon peak due
to the $e$-ph coupling.
Meanwhile, because of the Fano effect, this phonon structure is modified
by the impurity band to have a shoulder-like shape, rather than a $\delta$-like
form as in the localized electron case of Ref. \cite{ma00}.

In Fig. 4, we show the spectra of a system with a strong $e$-ph coupling of
$S$=0.50 eV, and a low doping rate of 3.12\%.
Comparing with Fig. 3(a), one clearly finds that, a large $e$-ph coupling
at the doped sites greatly contributes to the expansion of impurity band,
so that the Fermi edge is clearly observed in the whole system PES (full line).
Moreover, the second phonon side peak also appears due to the multi-phonon
processes aroused by the strong $e$-ph coupling, corresponding to a
step-like structure in the whole system PES, as affirmed in the inset.

\section{Summary}

We develop a new path-integral theory to calculate the PES of correlated
many-electron systems.
The isotopic shift in the ARPES of Bi2212 is investigated by this theory
based on a model including both $e$-$e$ and off-diagonal quadratic $e$-ph
interactions.
Our calculation demonstrates that the band shift is primarily triggered
by the $e$-ph coupling, while the presence of $e$-$e$ repulsion tends
to suppress this effect.
We also apply this theory to the MIH model on a doped cubic lattice to
clarify the spectral properties of BDD.
It is clearly shown in the PES that a semiconductor-metal phase transition
takes place due to the increases of $e$-ph coupling and dopant concentration.
Furthermore, the presence of Fermi edge and phonon step-like structure
indicates the co-existence of two basic characters of electron, itineracy
and localization, in BDD.

\end{document}